\documentclass{article}

\usepackage{pdfpages}
\usepackage{caption}
\usepackage{subcaption}
\usepackage{savesym}
\usepackage{hyperref}
\usepackage{lineno}
\usepackage[sort&compress]{natbib}

\usepackage{graphicx} 
\usepackage{gensymb}
\modulolinenumbers[1]

\usepackage{booktabs}
\savesymbol{splitbox}
\usepackage{adjustbox}
\usepackage[nomarkers, nolists,figuresonly]{endfloat}

\title{The main asteroid belt: the primary source of debris on comet-like orbits}

\graphicspath{{./}{figs/}}
\author
{P.M. Shober$^{1*}$, E.K. Sansom$^{1}$, P.A. Bland$^{1}$, H.A.R. Devillepoix$^{1}$, \and M.C. Towner$^{1}$, M. Cup\'ak$^{1}$, R.M. Howie$^{1}$, B.A.D. Hartig$^{1}$, S.L. Anderson$^{1}$\\
\\
\normalsize{$^{1}$Space Science \& Technology Centre, School of Earth and Planetary Sciences, Curtin University,}\\
\normalsize{GPO Box U1987, Perth, Western Australia 6845, Australia}\\
\normalsize{$^\ast$ E-mail: patrick.shober@postgrad.curtin.edu.au}
}

\begin{document}

\maketitle 


\baselineskip24pt


\begin{abstract}
Jupiter family comets contribute a significant amount of debris to near-Earth space. However, telescopic observations of these objects seem to suggest they have short physical lifetimes. If this is true, the material generated will also be short-lived, but fireball observation networks still detect material on cometary orbits. This study examines centimeter-meter scale sporadic meteoroids detected by the Desert Fireball Network from 2014-2020 originating from Jupiter family comet-like orbits. Analyzing each event's dynamic history and physical characteristics, we confidently determined whether they originated from the main asteroid belt or the trans-Neptunian region. Our results indicate that $<4\%$ of sporadic meteoroids on JFC-like orbits are genetically cometary. This observation is statistically significant and shows that cometary material is too friable to survive in near-Earth space. Even when considering shower contributions, meteoroids on JFC-like orbits are primarily from the main-belt. Thus, the presence of genuine cometary meteorites in terrestrial collections is highly unlikely.
\end{abstract}
  

\section{Introduction}
Jupiter family comets (JFCs) are short-period comets near the ecliptic plane that are strongly influenced by Jupiter. They originate from the outer solar system in the scattered disk \citep{duncan1997disk} and are comprised of primitive volatile-rich material. They are conventionally defined by their Tisserand's parameter ($2<T_{J}<3$), which is an approximately conserved value in the circular restricted three-body problem (object - Jupiter - Sun). This is a better way to classify the orbits of small planetary bodies compared to the traditionally used orbital periods \citep{levison1994long}. The physical properties of JFCs have been explored in numerous studies based on meteor/fireball analysis \citep{jenniskens2006meteor,borovivcka2007atmospheric,madiedo2014trajectory}, telescopic observations \citep{fernandez2005albedos,fernandez2013thermal}, dynamical modeling \citep{duncan2004dynamical,disisto2009population,nesvorny2010cometary,tancredi2014criterion}, and by space missions \citep{brownlee2004surface,ahearn2005deep,sunshine2006exposed,fornasier2015spectrophotometric}. Each method provides information about different size-ranges of objects on JFC-like orbits. 

When observing meteors streak across the night sky, we are typically witnessing the size-range of the smallest objects in the near-Earth space (dust-sized; $\ll10^{-3}$\,kg). Conversely, telescopic observations are instrumental in characterizing the most massive objects (hundreds of meters to kilometers in diameter). This gap between the meteor and telescopic observations is the primary range for the progenitors of meteorites. Usually spanning centimeters to meters in scale, this size-range has been interpreted from fireballs and the meteorites they produce on the ground \citep{borovivcka2015small,granvik2018identification}.

Based on telescopic observations and dynamical studies, many researchers believe there is a paucity of JFCs at sub-kilometer scales \citep{meech2004comet,fernandez2006population,nesvorny2017origin}. This lack of objects at smaller sizes is thought to be related to the inferred brief physical lifetimes ($10^{3}-10^{4}$\,yrs) of JFCs in the inner solar system \citep{levison1997kuiper,fernandez2002there,disisto2009population}. Due to their low-bulk density and high volatile content, JFCs should fade away relatively quickly when within $\sim$3\,au.


Nevertheless, despite the predicted sub-kilometer paucity, fireball observation networks have many accounts of durable centimeter to meter-sized meteoroids originating from JFC-like orbits \citep{brown2000fall,borovivcka2013kovsice,spurny2013trajectory,spurny2017ejby}. By examining meter-size terrestrial impactors, it was found that about $5-10\%$ had JFC-like orbits, relatively consistent with the flux estimates based on larger kilometer-scale near-Earth objects (NEOs) \citep{brown2016orbital}. Yet, they only found half of these displayed weaker than average structures based on their atmospheric ablation characteristics. Two of the five JFC-like events discussed were calibrated meteorite falls (Maribo and Sutter's Mill) likely to be from the main asteroid belt. Meanwhile, the remaining three events all came from US Government sensor (USG) data, two with a semi-major axis of $\sim2.9$\,au, i.e., well within the bounds of the outer main-belt. Also, it has since been demonstrated that the orbital information collected by USG events is generally unreliable \citep{devillepoix2019observation}. 

Additionally, another study recalculated the orbits for 25 meteorite falls and identified their likely source regions using an advanced NEO model \citep{granvik2018identification,granvik2018debiased}. The only meteorite that had a mean JFC probability $\geq50\%$ was an H5/6 ordinary chondrite, Ejby. Furthermore, the fall with the second-highest JFC source region probability was another H5, Ko\v{s}ice. However, both of these falls have relatively significant initial velocity uncertainties, increasing the uncertainty for the source region analysis. Carbonaceous-chondrites Maribo and Sutter`s Mill (both CM2), also had probabilities of coming from the JFCs above $20\%$. Nevertheless, no precisely observed meteorite falls have yet come from an unambiguously cometary orbit. 

The Taurid complex has also been discussed as being capable of larger, possibly meteorite-dropping, meteoroids \citep{brown2013meteorites}. These meteoroids can be hundreds of kilograms in size, an order of magnitude higher than other showers. Although other studies have found that while the Taurids can produce very large meteoroids, the Taurids are still very weak and were unlikely to produce meteorites \citep{borovivcka2017january}. Given the highly evolved orbit of comet 2P/Encke and the other non-cometary possible parent bodies in similar orbits, this subset of Taurids observed also does not likely reflect the vast majority of JFC meteoroids \citep{asher1993asteroids}. 

Based on debiased NEO orbital and absolute-magnitude distributions, it is expected that the contribution from the JFC region is a few percent on average \citep{bottke2002debiased,granvik2018debiased}. The contribution seems to increase at smaller sizes, reaching $\sim10\%$ of the NEO population below diameters of 100\,m \citep{granvik2018debiased}. However, this estimate assumes that JFCs are more likely to become dormant than disintegrate as JFC disruptions are less commonly observed compared to long-period comets. If this assumption breaks down at smaller sizes, then the JFC contribution to the NEO meteoroid flux may be negligible.

If the hypothesis that the physical lifetimes of sub-kilometer comets are very short is accurate, why do fireball networks still observe centimeter-meter debris originating from comet-like orbits? Does the larger cometary debris we observe impacting the Earth originate from comets? 

The massive orbital dataset collected by the Desert Fireball Network (DFN) was utilized to answer these questions. The DFN is the largest single photographic fireball network in the world, covering over $2.5$ million km$^{2}$ of Australia. This massive project is part of a worldwide collaboration, the Global Fireball Observatory (\url{https://gfo.rocks/}) (GFO), currently consisting of 18 partner institutions \citep{devillepoix2020global}. The GFO has a semi-automated data processing pipeline which sorts images, detects fireball events, and triangulates these events \citep{howie2017build,jansen2019comparing,towner2019fireball,sansom20193d,sansom2019determining}. 

\section{Materials and Methods}

\subsection{Experimental Design}
The dataset used for this study was collected by the Desert Fireball Network (DFN). Covering over one third of Australian outback, the DFN is the largest photographic fireball network globally. The observations made by the DFN are invaluable, and provide key insights into the debris impacting the Earth daily. 

When parsing through the dataset for this study, all events that originated from an orbit with a JFC-like Tisserand's parameter ($2<T_{J}<3$) and had a significant initial mass ($\geq0.01$\,kg) were gathered. The Tisserand's parameter is an approximately conserved value in the three-body problem and is regularly used to distinguish between different kinds of small planetary bodies \citep{levison1994long}. Typically this three body problem includes the sun, Jupiter, and the asteroid or comet; since Jupiter is seen as the primary perturber the parameter is written as $T_{J}$. 

\subsubsection{Trajectory Analysis \& Orbit Determination}
The fireball events detected by the DFN have atmospheric trajectories fitted using a modified straight-line least-squares (SLLS) and an extended Kalman smoother for the velocity profile \citep{sansom2015novel}. The uncertainties associated with the fitted fireball trajectories are propagated from the residuals of the fit itself along with the timing and positional uncertainties of the events` observations. The observational uncertainties are handled and incorporated by the Kalman filter.  The initial masses and corresponding uncertainties were also determined using a dynamic model in a reverse extended Kalman filter \citep{sansom2015novel}. The pre-atmospheric orbits are then determined numerically including any relevant perturbations by integrating the meteoroid`s state until it was outside the Earth`s sphere of influence \citep{jansen2019comparing}. The orbital uncertainties are then obtained with a Monte Carlo approach, by numerically integrating samples randomly drawn from within the initial state uncertainties at the top of the atmosphere.

\subsubsection{Addressing Observational Biases}
The DFN uses photographic observations of fireballs to help better understand the debris in the inner solar system. However, to obtain meaningful results from this dataset, we must first address any observational biases. 

The first bias we address is that the DFN is optimized for meteorite-dropping fireball events ($\sim0.5$ limiting magnitude) \citep{howie2017build}. In this study, we limit the meteoroids considered to a minimum initial mass of 10\,grams. This bias does not affect the results as both asteroidal and cometary populations are known to produce material within this size-range \citep{boehnhardt2004split,fernandez2009s}. 

Secondly, we must ensure that the DFN observations do not bias against either observing asteroidal or cometary debris due to the orbits they exist. The primary bias to be concerned with is due gravitational focusing. Since meteoroids with a lower relative velocity are more gravitationally focused, the terrestrial impact population contains a larger proportion of asteroidal impactors from orbits with smaller semi-major axis values. However, this study is concerned with meteoroids all originating from similar orbits, thus they are all weighted nearly equally. Other observational biases are associated with photographic fireball observations, however, none of these would have any affect on the results of this study. For a more exhaustive list and discussion of DFN observational biases please reference \citet{shober2020using}.


\subsubsection{Data Selection}
A manual check of all the fireball observations and triangulations, any events with poor data were removed (e.g. fireball was very far from cameras, convergence angles for triangulations were too small, etc.). The orbital uncertainties of the meteoroids must be sufficiently low to obtain meaningful statistics from the Monte Carlo numerical integrations. If the uncertainties are too large, the number of samples required to get statistically reasonable results may be computationally unfeasible. Additionally, meteoroids with $T_{J}$ values within three standard deviations of $T_{J}=3.0$ or $T_{J}=2.0$ were excluded to remove ambiguous events. 

In this study, only fireballs with no shower-associations were considered, i.e., sporadic. Therefore, any events with distinctly cometary physical and dynamical characteristics could indicate a process to preserve cometary material in this range for longer periods. Using primarily the $D_{J}$ similarity criterion, we sorted through associations with a limiting value of $D_{J} < 0.15$ \citep{jopek1993remarks}. A more inclusive limiting value was used to eliminate any potential shower events. We verified associations from this subset by comparing orbital elements, shower dates, right ascensions and declinations, initial velocities, nodal directions, arguments of perihelion, along with other similarity criterions \citep{southworth1963statistics,drummond1981test}. The meteor shower data used was taken from all showers listed by the IAU Meteor Data Center \footnote{\url{https://www.ta3.sk/IAUC22DB/MDC2007/}}, with a higher weight placed on established showers. 

The fireball data used within this study, including uncertainties, are available in the Appendix.

We characterized the physical and dynamical characteristics of the sporadic meteoroids on JFC-like orbits using fireball data from the DFN with the following goals: 
\begin{enumerate}
    \item Determine proportion of genetically cometary material in excess of 10\,g on sporadic JFC-like orbits  
    \item Assess whether or not the physical lifetimes of JFC meteoroids in the meteorite-dropping size range are shorter than the disassociation lifetimes of cometary streams  
    \item Find out if it is possible to get a meteorite from a JFC-orbit, and what it would look like.  
\end{enumerate}

\subsection{Dynamic Analysis}
\begin{figure}
    \centering 
     \begin{subfigure}[b]{0.49\textwidth}
         \centering
         \includegraphics[width=\textwidth]{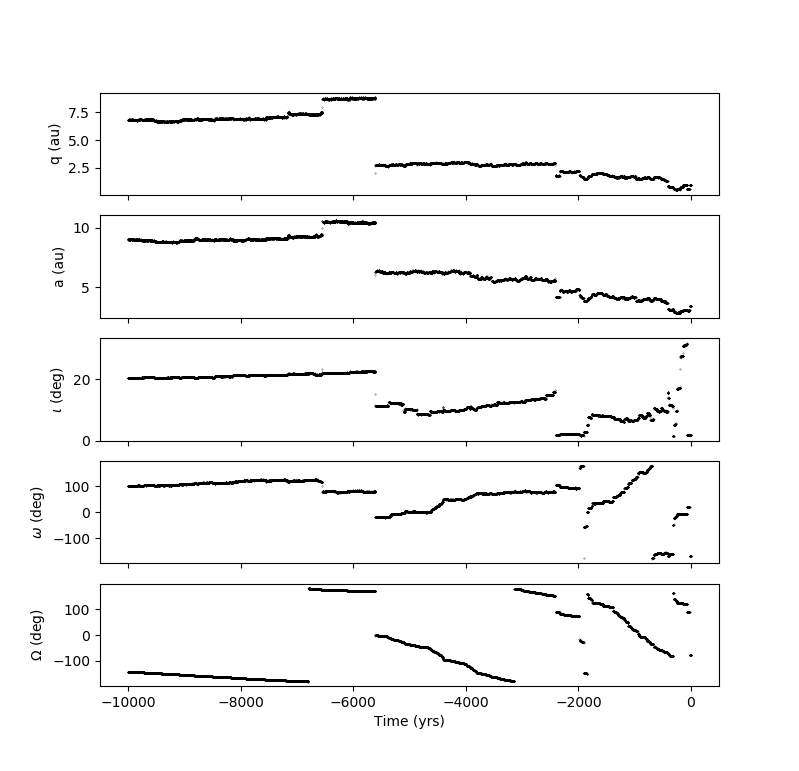}
         \caption{Unstable}
         \label{fig:unstable}
     \end{subfigure}
     \hfill
     \begin{subfigure}[b]{0.49\textwidth}
         \centering
         \includegraphics[width=\textwidth]{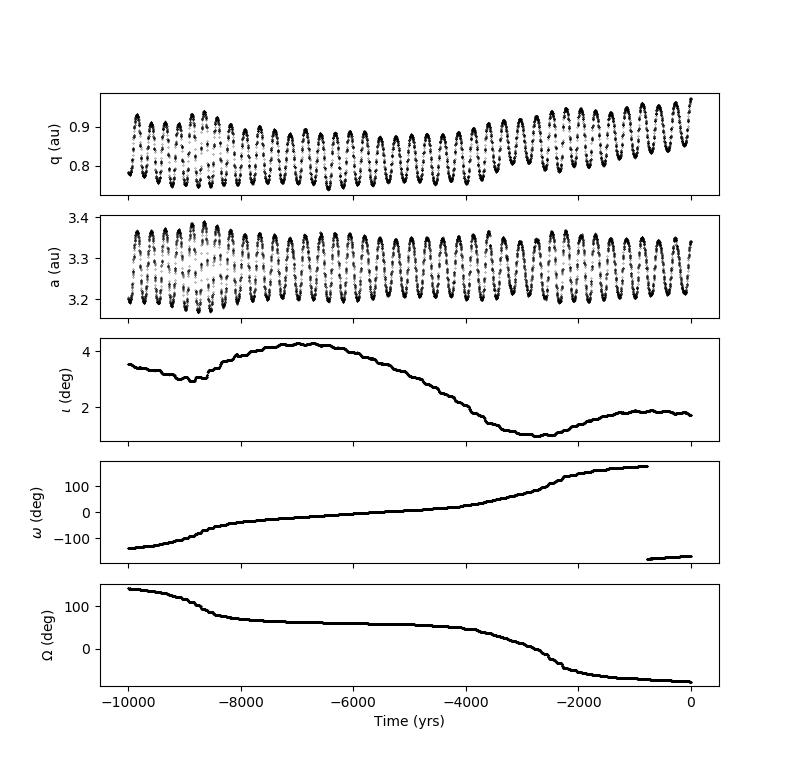}
         \caption{Stable}
         \label{fig:stable}
     \end{subfigure}
    \caption{Backward integrations -10\,kyrs of two particles generated within the triangulated trajectory uncertainties of event DN200104\_01 where q is perihelion, a is semi-major axis, $\iota$ is inclination, $\omega$ is argument of perihelion, and $\Omega$ is longitude of ascending node. On the left, a particle on a chaotic unstable orbit typical of a comet with numerous Jupiter close encounters. On the right, another particle provides an example of a stable orbit following a predictable path with no close encounters with Jupiter. This specific fireball event lasted over 5 sec and was recorded by 4 DFN stations, providing an excellent trajectory fit with very minimal uncertainties. Thus, this demonstrates the difficulty associated with backward propagation of meteoroids coming from orbits on the edge of two source regions (JFC and main-belt in this case).} 
    \label{fig:stable_vs_unstable} 
\end{figure}

As shown in previous studies, genuine JFCs originating from the outer solar system tend to move on chaotic orbits over thousands of years, encountering Jupiter frequently over their lifetime \citep{tancredi1995dynamical,levison1997kuiper}. In this work, we have adopted a similar search strategy to past studies by numerically integrating the fireball trajectories backward 10\,kyrs, identifying fireballs originating from `stable' and `unstable' orbits \citep{tancredi2014criterion,fernandez2014assessing,fernandez2015jupiter}. For each event, a Monte Carlo simulation was undertaken where ten simulations with 100~particles were integrated -10\,kyrs. These simulations were conducted using the IAS15 integrator, taking into account all planetary perturbations and close encounters considering all of the planets and the Moon \citep{2012A&A...537A.128R,2015MNRAS.446.1424R}. The particles were initialized within the formal triangulation uncertainties assuming a Gaussian distribution. The particles' histories were then each assessed and labeled as either `stable' or `unstable' over the previous 10\,kyrs. Those with frequent unpredictable jumps in their orbital elements were considered unstable (Fig. \ref{fig:stable_vs_unstable}\protect\subref{fig:unstable}), whereas those in resonances and followed smooth trajectories with little changes in perihelion distance were considered stable (Fig. \ref{fig:stable_vs_unstable}\protect\subref{fig:stable}). The final instability probabilities and associated uncertainties were then obtained by taking the mean and standard deviation of the sample means.

\subsection{Physical Analysis}
In order to characterize the physical properties, we determined the PE value for each fireball. The PE criterion was first introduced as a way to discriminate between different meteoroid compositions based on the ability to penetrate the atmosphere \citep{ceplecha1976fireball}. It has since been used in numerous studies to assess the physical nature of impacting meteoroids \citep{ceplecha1994impacts,brown2013meteorites,borovivcka2015small,brown2016orbital}. The PE value is given by: 
\begin{equation}
    PE = \log(\rho_{E}) - 0.42\log(m_{\infty}) + 1.49\log(V_{\infty}) - 1.29\log(\cos Z_{R})  
\end{equation}
where $\rho_{E}$ is the density of the atmosphere at the end the luminous trajectory (and the reason for the acronym `PE'), $m_{\infty}$ is the initial mass of the meteoroid in grams, $V_{\infty}$ is the initial velocity of the meteoroid in $km/s$, and $Z_{R}$ is the local entry angle measured from the zenith.

The PE values are divided into four types: ordinary chondrite-like (type I), carbonaceous (type II), short-period cometary (type IIIa), and weak cometary (type IIIb) \citep{ceplecha1998meteor}. The boundaries for these groups are denoted in Table~\ref{tab:pe_types}. The PE criterion is an extremely useful metric to evaluate the physical properties of meteoroids, however, the types are not strictly correct. While providing generally accurate results of relative strength, the physical composition of any individual meteoroid is uncertain due to the other factors that affect the durability while transiting the atmosphere (e.g. macro-scale fractures) \citep{popova2011very,borovivcka2020two}. 

We additionally computed the ballistic coefficient ($\alpha$) and the mass-loss parameter ($\beta$) based on the fireball observations \citep{gritsevich2006extra,gritsevich2007og,lyytinen2016implications}. Using these parameters one can quickly identify potential meteorite-dropping events \citep{sansom2019determining}. They can be calculated for any event showing some level of deceleration by using the velocity and height data (\url{https://github.com/desertfireballnetwork/alpha_beta_modules}). 

In Fig.~\ref{fig:lna_lnb}, we determined potential meteorite dropping fireballs using the $\alpha$-$\beta$ values. Assuming a 50\,g final mass is the minimum meteorite dropping mass, the meteorite dropping lines were calculated with the following equations: 
\begin{equation}
    \ln\beta = \ln[13.2 - 3 \ln(\alpha \sin\gamma)], \quad \mu  =  0 
    \label{eq:nospin_ordinary}
\end{equation}
\begin{equation}
    \ln\beta = \ln[4.4 - \ln(\alpha \sin\gamma)], \quad \mu  =  \frac{2}{3} 
    \label{eq:spin_ordinary}
\end{equation}
when assuming $\rho_{m}$ = 3500\,kg\,m$^{-3}$ (meteoroid density) and c$_d$\,A = 1.5 (product of drag coefficient and initial shape coefficient). Considering that the contribution from the outer main-belt may be significant, where many objects have low-albedos \citep{takir2015toward,demeo2014solar,demeo2015compositional} (e.g., C- and D-types), we also plotted the equivalent lines for a CM-like density \citep{consolmagno2008significance} ($\rho_{m}$ = 2240\,kg\,m$^{-3}$) using the following equations:  
\begin{equation}
    \ln\beta = \ln[14.09 - 3 \ln(\alpha \sin\gamma)], \quad \mu  =  0 
    \label{eq:nospin_CM}
\end{equation}
\begin{equation}
    \ln\beta = \ln[4.7 - \ln(\alpha \sin\gamma)], \quad \mu  =  \frac{2}{3} 
    \label{eq:spin_CM}
\end{equation}

Together, the dynamic simulations and analysis of the material properties can better elucidate what types of material regularly impact the Earth from JFC-like orbits in the considered size-range. 

\subsection{Statistical Analysis}
Based on the dynamical and physical analysis, the source populations can be inferred for each meteoroid in this study. However, given the sample size, what does this tell us about the entire population of small JFC-like meteoroids? 

In order to answer this question, we utilize a Markov Chain Monte Carlo (MCMC) to sample from the posterior probability distribution. The Bayesian model was created with the Python PyMC3 probabilistic programming package using the No-UTurn Sampler \citep{salvatier2016probabilistic}. A beta distribution was used as a prior with the hyperparameters ($\alpha$) based on a study that examined the dynamical characteristics of 58 near-Earth JFCs ($q<1.3$\,au) \citep{fernandez2015jupiter}. This corresponded to hyperparameters $\alpha=[1.38, 8.62]$ and an observation vector $c=[48, 2]$; the hyperparameters were weighted less as they were based on kilometer-scale objects. Additionally, as a comparison, an uninformed uniform prior of $\alpha=[1.0,1.0]$ was included. In total, fifteen thousand samples from the posterior were taken to estimate the posterior. To be counted as genetically JFC in origin in the model, the meteoroids would need to have a $>50\%$ probability of originating from an unstable orbit and be of PE types II or III. We considered a PE type II as plausibly cometary due to previous arguments that meteorites from JFCs could possibly be represented by primitive carbonaceous chondrites \citep{gounelle2008meteorites}. 

\section{Results}
In this study, we examined fireballs not associated with meteor showers detected by the DFN during four years of observations. All the fireballs considered were generated by meteoroids that originated from JFC-like orbits ($2<T_{J}<3$). The source region of these meteoroids, main-belt or JFC, was determined by analyzing the orbital stability and physical durability during atmospheric transit. It takes only $\sim1000$\,yrs to become orbitally disassociated from a JFC shower. Therefore, the identification of sporadic JFC material would indicate a minimum physical lifetime equal to the disassociation time \citep{tancredi1995dynamical}. However, a lack of sporadic cometary debris in fireball observations would be clear evidence that the physical lifetimes are less than one thousand years. 

\subsection{Orbital characteristics of sporadic JFC-like meteoroids}
JFCs move on chaotic orbits compared to main-belt asteroids; on the order of thousands of years \citep{tancredi1995dynamical}. During their lifetimes, their orbital evolution is controlled by the numerous close encounters they have with Jupiter. Of the 50 sporadic JFC-like fireball events analyzed in this study, nearly all the meteoroids come from stable orbits over the previous 10\,kyrs (see Appendix). This is inconsistent with the dynamics diagnostic of JFCs. 

\begin{figure}
    \centering
    \includegraphics[width=0.9\columnwidth]{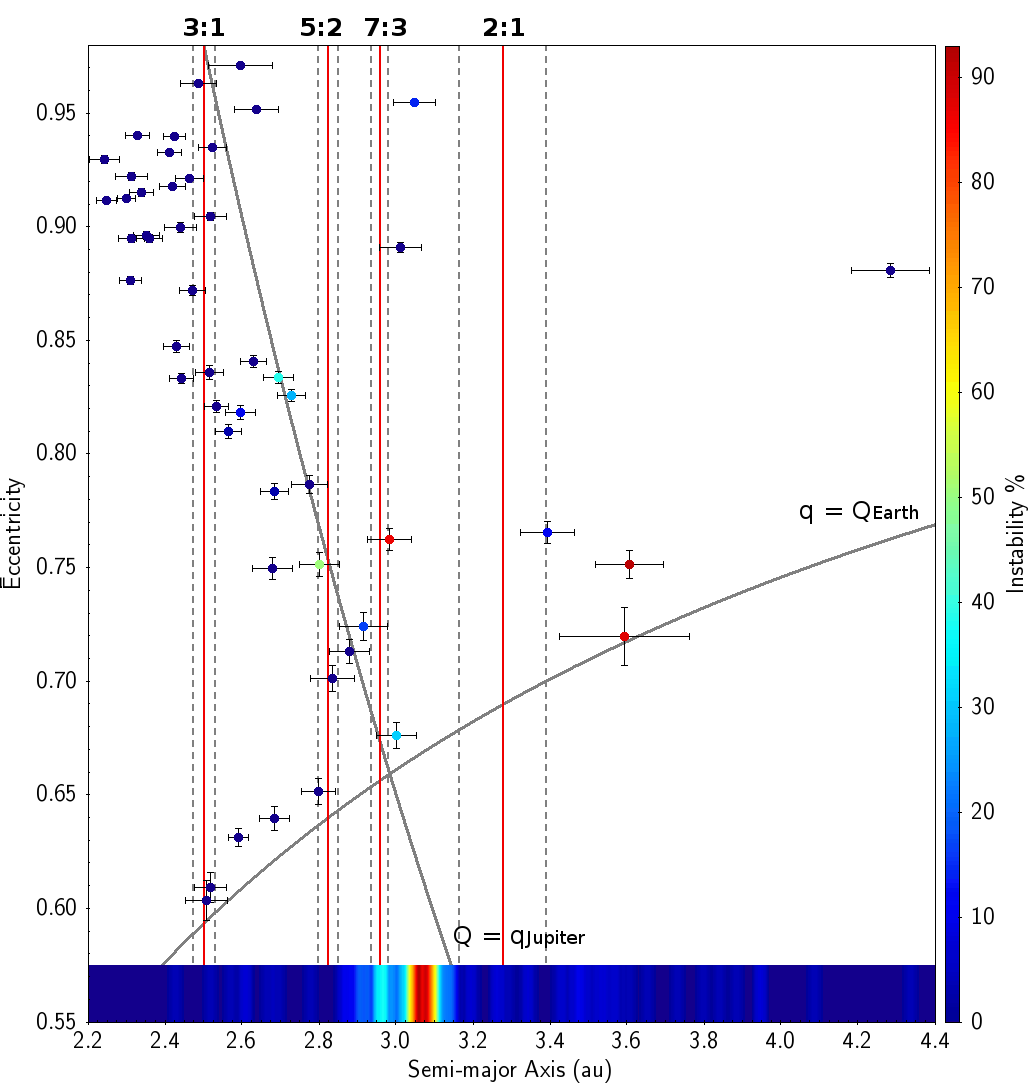}
    \caption{Triangulated semi-major axis vs. eccentricity for 50 non-shower JFC-like fireballs observed by the DFN that have predicted initial masses greater than 10\,g. Error bars denote the formal $1\sigma$ orbital uncertainties associated with each meteoroid, and coloration is indicative of the orbital stability determined through numerical integrations -10\,kyrs. The vertical red lines indicate the major MMRs with the dashed gray lines marking the maximum libration in semi-major axis \citep{tancredi2014criterion}. Most of the events originate from stable orbits near the MMRs. The densogram (\protect\url{http://www.star.bris.ac.uk/~mbt/stilts/sun256/layer-densogram.html}) at the bottom of the plot represents the semi-major axis density for near-Earth ($q<1.3$\,au) JFCs within according to the NASA HORIZONS system (\protect\url{http://ssd.jpl.nasa.gov/horizons.cgi}).}
    \label{fig:ae}
\end{figure}

Only three events have a high probabilities of originating from an unstable orbits. However, one of these comes from within 3\,au and crosses one of the major mean-motion resonances (MMRs). Events DN150817\_01 and DN161028\_02 have the highest probability of originating from unstable cometary orbits, with $86.1\pm5.5\%$ and $90.8\pm3.0\%$ of the particles integrated being chaotic respectively. As seen in Fig.~\ref{fig:ae}, event DN150816\_04 overlaps the 7:3\,MMR when considering triangulation uncertainties, whereas, events DN150817\_01 and DN161028\_02 both lie outside the normal range of the main-belt with a semi-major axis of $\sim3.6$\,au. Several meteoroids have inconclusive dynamical histories; these include six events where several of the particles integrated evolved chaotically over thousands of years. On the other hand, nearly all of these inconclusive events come from orbits near MMRs, thus indicating that orbital uncertainty is very likely the cause of the inconclusive results. Otherwise, $>80\%$ of the events originate from extremely stable orbits nearby one of the primary MMRs (Fig.~\ref{fig:ae}). 


 A large majority of the fireballs examined in this study do not fall within the range consistent with observed kilometer-scale JFCs. As seen in Fig.~\ref{fig:ae}, a majority of the near-Earth JFCs exist near $\sim3.1$\,au, where only a handful of events originate. However, only one of the four events in this range comes from an unstable orbit. This is in line with a previous model, where they found that main-belt material can contaminate the JFC population but would primarily be contained to the original bounds of the main-belt \citep{hsieh2016potential}. Albeit, four of the 50\,events examined fall beyond the 2:1\,MMR, and of these, only two have higher probabilities of originating from unstable orbits (Fig.~\ref{fig:ae}). Additionally, the inclinations of the fireballs examined tend to be relatively diffuse, but are most concentrated towards low-values within the $0-25^{\circ}$ range. This is more reconcilable with the main-belt inclination distribution than the JFC distribution, which is more highly concentrated around $\sim12^{\circ}$ (see Appendix).

\subsection{Physical characteristics of sporadic JFC-like meteoroids}
 The PE criterion characterizes the composition of meteoroids based on their ability to penetrate the atmosphere \citep{ceplecha1976fireball}. This metric is empirically determined based on the terminal height of the fireball, the entry mass, the velocity at the top of the atmosphere, and the impact angle. The PE criterion has been used for decades to understand the physical properties of meteoroids based on their atmospheric ablation characteristics \citep{borovivcka2015small,brown2016orbital}. The values are traditionally split accordingly \citep{ceplecha1998meteor}: 
\begin{table}[h!]
    \centering
    \begin{tabular}{| l | c c |}
         \hline
         type I     & $PE > -4.60$            & Ordinary chondrite-like \\
         type II    & $-5.25 < PE \leq -4.60$ & Carbonaceous chondrite  \\
         type IIIa  & $-5.70 < PE \leq -5.25$ & Short-period cometary   \\
         type IIIb  & $PE \leq -5.70$         & Weak cometary material  \\
         \hline
    \end{tabular}
    \caption{Traditional PE classifications based on the atmospheric density at terminal height \citep{ceplecha1976fireball}.}
    \label{tab:pe_types}
\end{table}

\begin{figure}
    \centering
    \includegraphics[width=1.0\columnwidth]{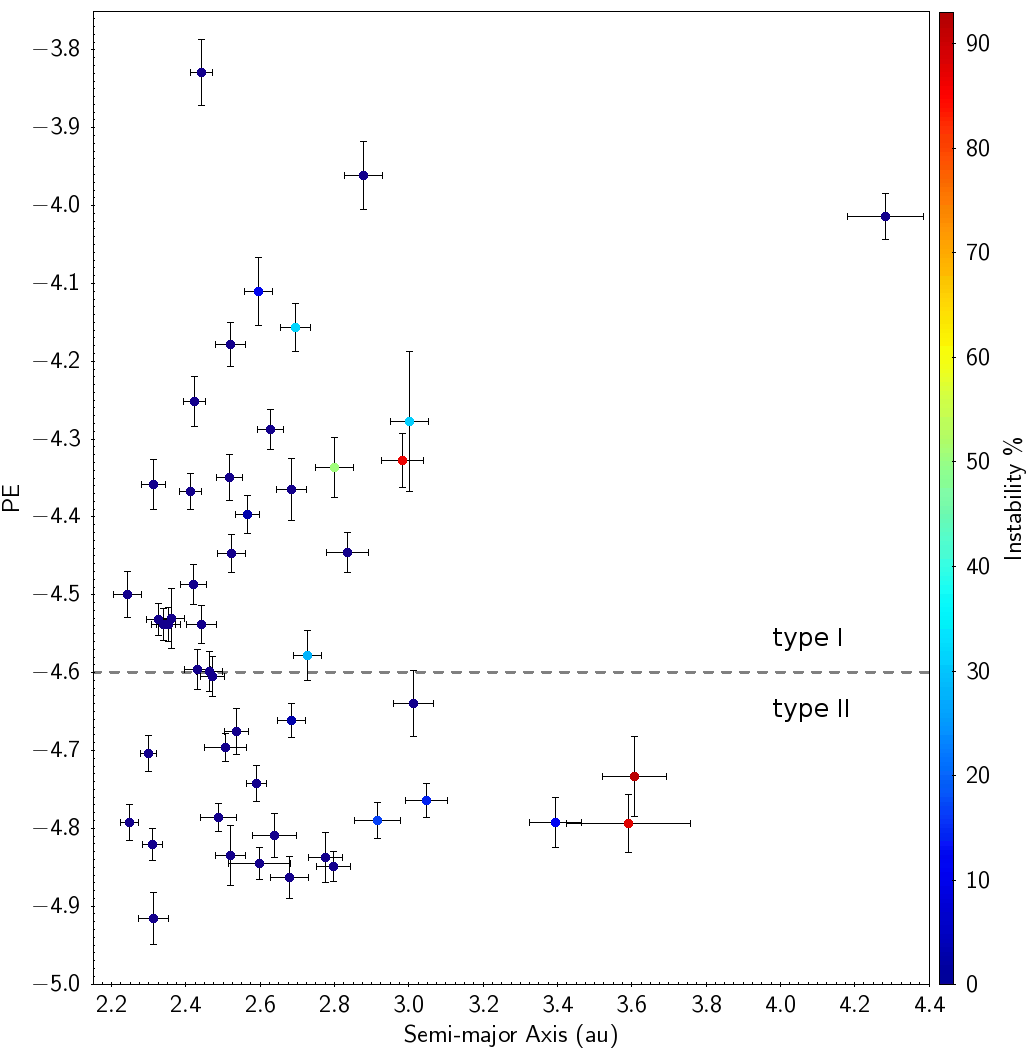}
    \caption{Sporadic JFC-like fireballs observed by the DFN with a minimum initial mass of 10\,g. The end heights of meteors can be used to ascertain the physical-nature of meteoroids using the PE criterion \citep{ceplecha1976fireball}. The PE values for the 50 fireballs examined are overwhelmingly stronger than the traditional comet types (Types IIIa and IIIb). The error bars associated with each meteoroid`s semi-major axis and PE value are due to triangulation uncertainties.}
    \label{fig:PE_a}
\end{figure}

The use of the PE criterion can be quite valuable when analyzing many fireballs; however, the types are not strictly exclusive and should not be taken as absolute truth. Other properties beyond physical composition can blur the lines between the types, such as macro-scale cracks \citep{popova2011very,borovivcka2020two}. As expected, considering the dynamic analysis, many of the JFC-like fireballs in this study are quite durable and strong (Fig.\ref{fig:PE_a}). We only see meteoroids of types I or II, i.e., none fall into the standard categories associated with cometary material. Out of the 50 fireballs, 28 would be classified as type I and 22 as type II. Some short-period showers actually can produce more durable debris with similar strengths \citep{brown2013meteorites}. Therefore, the PE criterion alone is not a sure way to distinguish between different materials for an individual fireball. In general, the PE types can give a sense of how strong the initial meteoroid was, but when combined with the dynamic analysis, the source population of a meteoroid can be more confidently obtained (main-belt or JFC). 

\begin{figure}
    \centering
    \includegraphics[width=1.0\columnwidth]{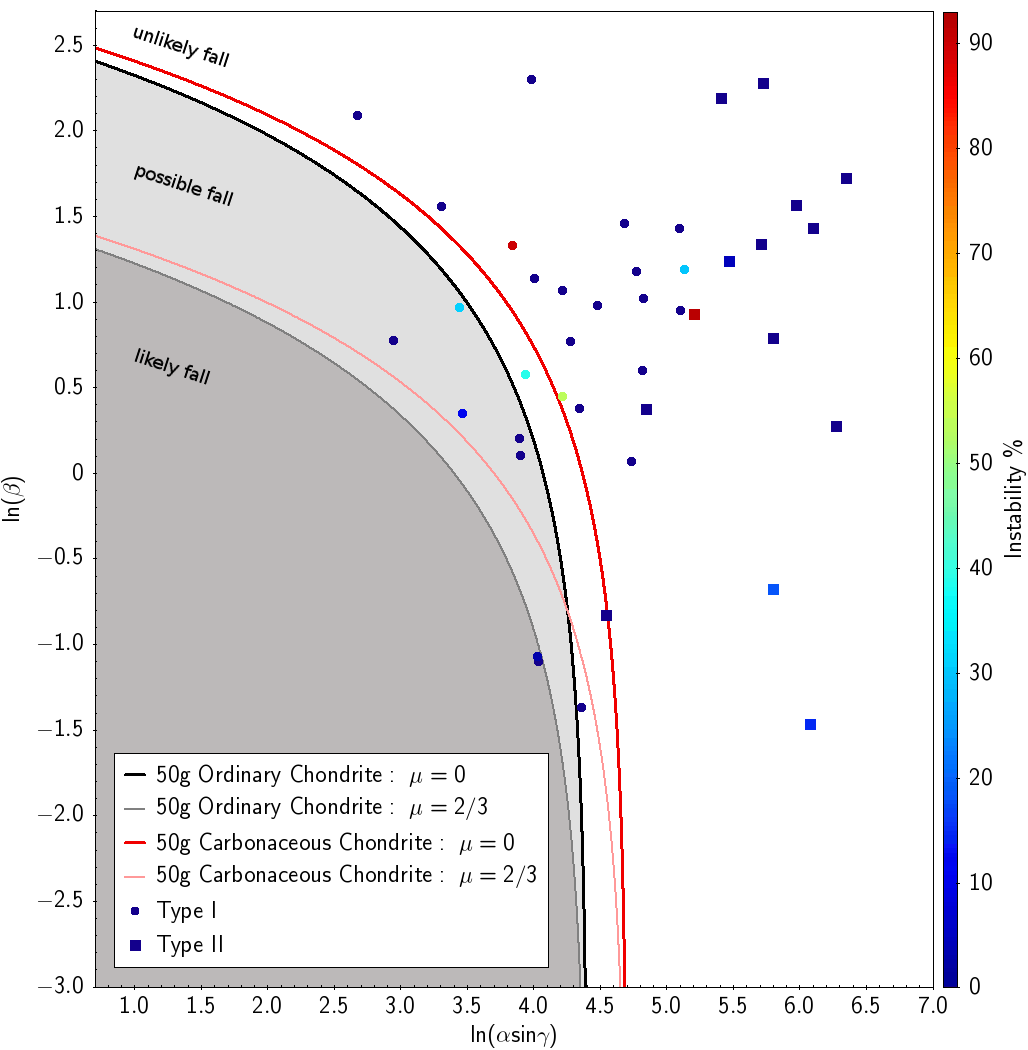}
    \caption{Distribution of non-shower JFC fireballs with enough deceleration to determine $\alpha - \beta$ values (at least $\sim20\%$ deceleration). $\gamma$ is the trajectory slope relative to the horizontal. If a macroscopic event is considered to have a final mass of $\geq50$\,g, assuming $\rho_{m}$ = [2240, 3500]\,kg\,m$^{-3}$ (carbonaceous and ordinary chondrite respectively) and c$_{dA}$ = 1.5, meteorite dropping events can easily be identified given the range of possible shape change coefficients ($\mu$).}
    \label{fig:lna_lnb}
\end{figure}

\subsubsection{Meteorite production from JFC-like meteoroids}
Traditionally, the empirical criteria to predict whether a fireball has produced a meteorite on the ground (``a fall'') is height and velocity below 35\,km and 10 kms$^{-1}$ respectively at the end of the luminous portion of the atmospheric trajectory \citep{wetherill1981fireballs}. This gives us an easy way to identify possible recoverable falls quickly. A more rigorous way to determine which fireballs result in meteorites has since been demonstrated, where the ballistic coefficient ($\alpha$) and the mass-loss parameter ($\beta$) are fit to the fireball observations \citep{sansom2019determining}. 

In Fig.~\ref{fig:lna_lnb}, we see that several of the JFC-like fireballs examined in this study are quite durable. At least some events are likely to have meteorites on the ground $>50$ g (DN150905\_01 and DN191020\_02). However, these events are likely asteroidal in origin based on their dynamic stability over the previous 10\,kyrs and their PE classification. No events with high instabilities over the previous 10\,kyrs or of type\,II drop meteorites in this study. Thus, meteorites from JFC-like orbits do occur, but they are all representative of main-belt material. 

\section{Discussion}
Based on dynamical and physical analysis of sporadic JFC-like ($2<T_{J}<3$) fireballs with initial masses $>$10\,g, we find a clear abundance of stronger dynamically stable meteoroids (Fig.~\ref{fig:PE_a}). As seen in Fig.~\ref{fig:ae}, most meteoroids evolve in a deterministic way over the previous 10\,kyrs, and have proximity to major MMRs. Dynamic instability is a diagnostic feature of JFCs originating from the outer solar system, as they suffer frequent close encounters with Jupiter \citep{tancredi1995dynamical,tancredi2014criterion}. The predictable evolution of nearly all the meteoroids in this study is therefore indicative of their likely origin in the main belt. Additionally, events with indeterminate histories, where a non-negligible subset of particles evolved chaotically, also mostly lie near the primary MMRs. As does one of the three events with $>80\%$ instability (Fig.~\ref{fig:PE_a}).
This proximity to MMRs could indicate possible main belt origins. Moreover, JFCs' arguments of perihelia tend to concentrate near $0^{\circ}$ and $180^{\circ}$ \citep{quinn1990planetary}, and there is no such correlation for the sporadic JFC-like fireballs in this study (see supplementary material). This further supports a dominantly main belt, asteroidal source.

Based on the calculated PE values, most of the events are also very durable and are classified as type I chondritic material. In fact, all of the fireballs in this sample are either classified as chondritic or carbonaceous chondritic material (type I or II; Fig.~\ref{fig:PE_a}). No events have PE types traditionally associated with cometary material (type III). However, it has been previously argued that primitive carbonaceous chondrites may have come from cometary sources \citep{gounelle2008meteorites}. Thus, any DFN events that likely evolved chaotically over the previous 10\,kyrs ($>50\%$) and are of type II were also considered as cometary. At most only 2 meteoroids studied here are likely to originate from a short-period comet (DN150817\_01 and DN161028\_02). However, this would require centimeter to meter cometary debris to be stronger than expected. The subset of weaker type II meteoroids in our data are also generally smaller ($<50\,g$), perhaps debris released by dark carbonaceous asteroids in the outer main-belt \citep{takir2015toward,demeo2014solar,demeo2015compositional}. A novel process has been recently observed to release centimeter-sized particles on carbonaceous asteroid Bennu during the OSIRIS-REx mission \citep{lauretta2019episodes}. 



Albeit, caution should be used with such a criterion to characterize meteoroids; a wide range of meteorite types have been recovered from type II/III events. For example, Almahatta Sitta (2008 TC3), Ko\v{s}ice, and Sutter`s Mill had PE values near the II/IIIa boundary, despite all being of different meteorite types on asteroidal orbits (polymict breccia, H5, and CM2 respectively) \citep{borovivcka2015small,brown2016orbital}. This is indicative that other physical factors, such as macroscopic cracks or porosity, can significantly affect the apparent strengths of impactors, and that there is still much unknown about the physical nature of meter-scale NEOs \citep{borovivcka2017january,borovivcka2020two}. The PE criterion is useful to characterize the relative strength of meteoroids, but how much the meteorite type influences the strength is debatable. 

\subsection{Meteoroids beyond the 2:1 resonance}
Previous studies have shown that a minor component of main-belt material can be transferred to JFC-like orbits, even in some circumstances, onto dynamically unstable orbits via close encounters with the terrestrial planets \citep{fernandez2002there,fernandez2014assessing,hsieh2016potential,shober2020did,shober2020using}. Nevertheless, these objects' semi-major axes are predicted to be primarily constrained to the original range of the main-belt \citep{hsieh2016potential}. Here we see that of the four events beyond 3.27\,au, two are likely asteroidal in origin based on their PE classifications and stability (Fig.~\ref{fig:PE_a}). The two most probable events to originate from unstable comet-like orbits are DN150817\_01 and DN161028\_02, with $86.1\pm5.5\%$ and $90.8\pm3.0\%$ probabilities of having chaotic histories based on Monte Carlo simulations along with orbits beyond the 2:1\,MMR (each around $a\sim3.6$\,au). The meteoroids from events DN150817\_01 and DN161028\_02 both fall into the type II, nominally carbonaceous and equally or more durable than many of the main-belt debris (Fig.~\ref{fig:PE_a}). Even if both events are genuinely cometary in origin, they only comprise $4\%$ of the sample. 

\subsection{Statistical Significance}
Let's consider the null hypothesis to be that the source regions of sporadic meteoroids on JFC orbits match those of kilometer-scale JFCs. A P-test was used to determine if our results were significant enough to reject the null hypothesis. The P-value reflects the probability of observing an as or more extreme result. We considered a P-value of 0.01 to be significant enough to reject the null hypothesis. At most, two events are genetically cometary within the dataset (DN150817\_01 and DN161028\_02). Both of these events meet the criteria of being relatively weaker (type II) and likely originating from unstable orbits over the previous 10\,kyrs ($>50\%$). We assume the probability of observing a meteoroid from the main-belt within our sample matches the maximum value found from a dynamical study based on telescopic observations (19 out of 58 JFCs from the main-belt) \citep{fernandez2015jupiter}. From these results, we find the P-value to be equal to $3.1 \times 10^{-21}$. This extraordinarily low probability clearly indicates we can reject the null hypothesis. Thus, the source region for centimeter-scale meteoroids on JFC-like orbits does not match the cometary kilometer-sized bodies.

We additionally utilized Bayesian inference to estimate the proportion of main-belt material on JFC-like orbits \citep{salvatier2016probabilistic}. The informed prior used is a beta distribution with hyperparameters $\alpha=[1.38, 8.62]$ and an observation vector $c=[48, 2]$. The hyperparameters are derived from observational data, but the weight of these prior observations on our expected values was slightly reduced as they corresponded to kilometer-scale objects \citep{fernandez2015jupiter}. Using a Markov-Chain Monte Carlo to draw samples from the posterior distribution, we found that sporadic JFC-like meteoroids in NEO space are $82.1\pm4.9\%$ from the main-belt and $17.9\pm4.9\%$ from the trans-Neptunian region with our reasonably informed prior. Whereas when an uninformed prior is used (i.e., the prior probability of observing a JFC or main-belt meteoroid is equal), the relative components are $94.2\pm3.2\%$ from the main-belt and $5.8\pm3.2\%$ from JFCs.

Furthermore, while processing the data in this study, we found that of the meteoroids originating from JFC-like orbits with initial masses greater than 10\,g, about $78\pm3\%$ were sporadic and $22\pm3\%$ were associated with showers. Thus, even when considering showers along with the sporadic population, it is likely that main-belt material makes up the majority of near-Earth material on orbits with $2<T_{J}<3$ within this size range. Evidently the Tisserand's parameter is not a very informative metric when analyzing meteoroids in the inner solar system.

\subsection{Why are there no sporadic cometary meteoroids?}

\subsubsection{Alternative explanations}
A possible explanation for the lack of dynamically unstable meteoroids in our sample could be that non-gravitational forces are decoupling the material from Jupiter. Previous work has found that sporadic micron-sized meteoroids of the zodiacal cloud are not able to decouple from Jupiter via planetary perturbations along with solar radiation pressure before having another close encounter removing them from Earth-crossing orbits \citep{nesvorny2010cometary}. However, the meteoroids studied by \citet{nesvorny2010cometary} are in a completely different size range compared to those discussed here, and the effects on these meteoroids are very different. 

Nevertheless, non-gravitational forces are highly unlikely to be responsible for the meteoroids with asteroidal dynamics in our data set. Non-gravitational forces, such as the Yarkovsky effect, would not be efficient enough to decouple these meteoroids before having another close encounter with Jupiter ($<10^{3}$\,yrs). 

Alternatively, one could argue that we may not expect to see many JFC meteoroids on these low-perihelion orbits or within the examined size range (10 grams to a few kilograms). Firstly, there have been multiple studies on Earth-crossing JFCs with the current number known to be $\sim140$\,objects (\url{http://ssd.jpl.nasa.gov/horizons.cgi}). A previous study investigated the dynamic stability of 58\,JFCs in orbits with $q<1.3$\,au, and they found that most of the objects (over two thirds) move on unstable orbits over $\pm10$\,kyrs \citep{fernandez2015jupiter}. Thus, showing at the kilometer scale objects on these orbits are dominated by genetically cometary material. Secondly, larger meteoroids (centimeter or larger) are capable of being produced by comets through a few mechanisms such as cometary splitting or gas drag \citep{boehnhardt2004split,fernandez2009s}. Thus, we should expect to observe cometary meteoroids within the range detected by the DFN. 

\subsubsection{Short physical-lifetimes}
Given the numerous JFCs on Earth-crossing orbits and their ability to produce meteoroids within the range considered in this study, it is surprising that we observe nearly no meteoroids clearly sourced from the JFC population \citep{fernandez2006population}. According to debiased NEO models, as much as $10\%$ of the NEO population at the smallest sizes considered ($\sim100$\,m), should be sourced from the JFC population \citep{granvik2018debiased}. Yet, in this sample, we see no distinctly cometary meteoroids. 

This lack of cometary debris within the sporadic population strongly supports the hypothesis that sub-kilometer JFCs have limited physical lifetimes in the inner solar system \citep{meech2004comet,fernandez2002there,fernandez2006population,nesvorny2017origin}. The physical lifetimes of centimeter-meter debris must be less than $\sim1000$\,years (the decoherence time). This physical breakdown is consistent with the observations that there are less high-volatile asteroids at low-perihelion distances \citep{granvik2016super}. Planetary debris seems to break down in the inner solar system closer to the sun, depending on the object's size and volatile content. At large sizes ($>10^{3-4}$\,m), material from the trans-Neptunian region is dominant on JFC-like orbits. However, as the diameters decrease to typical meteorite-producing objects, the physical lifetimes are likely so short that main-belt material becomes a more viable source. Conversely, according to \citet{nesvorny2010cometary}, micron--sized meteoroid impacts on Earth are dominated by JFC particles ($\sim85\%$ of the mass influx). If accurate, these results suggest that dust-sized and kilometer-sized objects on JFC-like orbits are genetically cometary, however, intermediate sizes are sourced from the main asteroid belt. This is consistent with the prediction by \citet{nesvorny2010cometary} that cm-sized particles ejected from comets likely quickly disrupt ($<10$\,kyrs) and form the robust micron-sized population observed within the zodiacal cloud. 

\subsection{Meteorites from Jupiter-family comets}
A couple of the fireballs likely resulted in surviving meteorites $>50$\,g (Fig.~\ref{fig:lna_lnb}), however, all of these are stable dynamically over the previous 10\,kyrs and only have PE values of type I. Thus, even in the occasion where we do have some genetically cometary material impact the Earth, according to our sample, it is unlikely to survive. It has been argued that primitive carbonaceous chondrites may be the best candidates to originate from cometary bodies, however, given our results, this may be only possible if the meteoroid was recently released from the parent body (within a thousand years) \citep{gounelle2008meteorites}. While the cosmic ray exposure ages are significantly shorter for some carbonaceous chondrites ($<1$\,Myr), they are still usually much longer than the $\sim10^{3}$\,yrs expected from our results \citep{eugster2006irradiation}. Therefore, meteorites from $2<T_{J}<3$ orbits are expected, but they will not be cometary. 

\subsection{Limitations}
This study is constrained by the amount of data collected by the DFN. However, given that the DFN is the largest fireball network in the world, this cannot be improved upon. Additionally, the PE value methodology is a very limited metric of meteoroid strength as macroscopic features of the meteoroids can significantly affect the strengths. However, the PE value was utilized liberally in this study to determine potentially cometary meteoroids. Any meteoroid with a slightly weaker strength was considered as potentially cometary. Even with this very inclusive criterion, we found nearly no meteoroids that fulfilled our physical and orbital requirements to be considered cometary. The results clearly demonstrate that cometary debris larger than dust-sized is too friable to survive in near-Earth space for longer periods. 


\section{Summary}
Sporadic fireball data collected by the DFN demonstrates there is a lack of cometary material on JFC-like orbits for meteoroids in the gram to kilogram size range. This supports the short physical lifetime hypothesis for JFC material \citep{meech2004comet,fernandez2002there,fernandez2006population,nesvorny2017origin}. Additionally, it indicates that the primary source region for Earth-crossing material on JFC-like orbits ($2<T_{J}<3$) is size-dependent. A majority of the objects at dust-sizes and on kilometer-scales are likely sourced from the trans-Neptunian region of the outer solar system \citep{nesvorny2010cometary,fernandez2015jupiter}. However, objects centimeters to meters in size are dominated by asteroidal material diffused out from the main-belt. This diffusion process likely occurs via some combination of orbital resonances, Kozai resonances, non-gravitational forces, and close encounters with terrestrial planets \citep{bottke2002debiased,fernandez2014assessing,hsieh2016potential,shober2020did,shober2020using}. In this study, we found a minor fraction of unstable possibly cometary objects, but the number of fireballs observed to originate from stable orbits from the main-belt over the previous 10\,kyrs is statistically significant. We estimate that as much as $94.2\pm3.2\%$ of Earth-crossing centimeter-sized debris on these orbits is sourced from the main-belt. Furthermore, considering that $22\pm3\%$ of the JFC-like fireballs observed by the DFN over the previous five years are associated with showers, our results suggest that the majority of centimeter material on JFC-like orbits in near-Earth space is main-belt in origin. Thus, we should expect to see ordinary types of meteorite falls coming from Jupiter-crossing orbits. 

\section{Acknowledgments}
This work was funded by the Australian Research Council as part of the Australian Discovery Project scheme (DP170102529). SSTC authors acknowledge institutional support from Curtin University. This work was also supported by resources provided by the Pawsey Supercomputing Centre with funding from the Australian Government and the Government of Western Australia. This research made use of TOPCAT for visualization and figures \citep{taylor2005topcat}. This research also made use of Astropy, a community-developed core Python package for Astronomy \citep{robitaille2013astropy}. Simulations in this paper made use of the REBOUND code which can be downloaded freely at http://github.com/hannorein/REBOUND \citep{2012A&A...537A.128R}. The authors declare no competing interests. All data needed to evaluate the conclusions in the paper are present in the paper and/or at \url{https://doi.org/10.5281/zenodo.4710556}.

\clearpage

\bibliography{references}

\end{document}